\documentclass[prb,twocolumn,superscriptaddress,showpacs]{revtex4}
\usepackage{amsmath,amssymb,graphicx,color,soul,bm,epstopdf}

\begin{document}

\title{Dissipative solitons in semiconductor microcavities at finite temperatures}

\author{D. V. Karpov}
\affiliation{Institute of Photonics, University of Eastern Finland, P.O.Box 111 Joensuu, FI-80101 Finland}
\affiliation{Department of Physics and Technology of Nanostructures, St. Petersburg Academic University, 8/3 Khlopina, St.-Petersburg, 194021 Russia}

\author{I. G. Savenko}
\affiliation{COMP Centre of Excellence at the Department of Applied Physics, Aalto University School of Science, P.O. Box 13500, FI-00076 Aalto, Finland}
\affiliation{National Research University of Information Technologies, Mechanics and Optics (ITMO University), St.-Petersburg, 197101, Russia}

\author{H. Flayac}
\affiliation{Institute of Theoretical Physics, Ecole Polytechnique F\'ed\'erale de Lausanne (EPFL), CH-1015 Lausanne, Switzerland}

\author{N. N. Rosanov}
\affiliation{National Research University of Information Technologies, Mechanics and Optics (ITMO University), Saint-Petersburg 197101, Russia}
\affiliation{Vavilov State Optical Institute, St.-Petersburg 199034, Russia}
\affiliation{Ioffe Physical Technical Institute, St.-Petersburg 194021, Russia}


\begin{abstract}
We consider exciton polaritons in a semiconductor microcavity with a saturable absorber in the growth direction of the heterostructure. This feature promotes additional nonlinear losses of the system with the emergence of bistability of the condensate particles number on the nonresonant (electrical or optical) excitation intensity. Further we demonstrate a new type of bright spatial dissipative exciton-polariton soliton which emerges in the equilibrium between the regions with different particle density.
We develop protocols of soliton creation and destruction. The switch to a soliton-like behavior occurs if the cavity is exposed by a short strong laser pulse with certain energy and duration. We estimate the characteristic times of soliton switch on and off and the time of return to the initial cycle. In particular, we demonstrate surprising narrowing of the spatial profile of the soliton and its vanishing at certain temperature due to interaction of the system with the thermal bath of acoustic phonons. We also address the role of polariton-polariton interaction (Kerr-like nonlinearity) on formation of dissipative solitons and show that the soliton may exist both in its presence and absence.
\end{abstract}
	
\maketitle


\section{Introduction} 
Saturable absorption is a widely used phenomenon in laser optics.~\cite{RefTrager, RefRosanovb1, RefStaliunas, RefAkhmedievb1, RefAkhmedievb2, RefLugiato2015} One important signature of it is nonlinear response of the system saturation on the input power increase. This effect is commonly used in the mode-locked solid state lasers~\cite{RefKeller, RefElsass, RefTaranenko, RefGenevet} aimed at producing extremely short light pulses. Heterostructures with embedded saturable absorber paved the way for studies of dissipative solitons (DSs) which became the focus of optics research about two decades ago~\cite{RefRosFed1992, RefStenger, RefRosKho1988, RefLuo, RefAkhmediev, SolitonPixels, SolitonsInMicrocavities, Sich, SwitchingPixels} due to their fundamental properties and potential for various applications in information processing~\cite{RefKuszelewicz, RefTlidi} Theoretical work on cavity solitons ~\cite{InteractionDS, Firth} has stimulated a variety of experiments.~\cite{Taranenko,Weiss} 
In heterostructure devices with saturable absorption, solitons can be engineered via geometry and alloy composition control.~\cite{RefKim,RefLiverini,RefMaas, RefSavenkoPSS} 

In this manuscript, we propose a new kind of cavity exciton--polariton-based DS and develop protocols aimed at its creation (switch on) and destruction (switch off). Exciton polaritons (later, polaritons) represent hybrid light-matter bosonic quasiparticles emerging in high-quality (high-Q) semiconductor microcavities.~\cite{RefImamoglu,BB,Racquet} They have half-photonic character allowing for fast propagation of the particle wavelets and also they represent half-excitons with nonlinear self-interaction. 
Polaritons have proven to be highly promising entities from both the fundamental and application-oriented points of view. Moreover, thanks to recent technological progress, high-Q microcavities of any geometry are routinely produced. Indeed, the state-of-the-art fabrication technology allows for creation of various semiconductor heterostructures with desired spatial patterns in the lateral directions. For instance, producing confining potentials of various kinds \cite{BB,Racquet,trans22,trans33,graphene} results in a growing number of theoretical proposals.~\cite{GaoPRB2013, LiewPRL2008, PavlovicPRL2009, EOArX2013, FlayacRouter}
Another reason why polaritons attract growing interest is that they can form quasi-Bose--Einstein condensation (BEC)~\cite{exit} which is similar to BEC of quasiparticles in other mesoscopic systems.~\cite{photon, Light,  MagnonsBEC, Vainio2015, IndexBEC, Fleisch}

{ {One more important reason why microcavities are advantageous over classical optical systems is the ability to access nonlinear Kerr-like media with large exciton-mediated response of the system due to particle self-scattering. This scattering leads to further reduction of the required input power and the characteristic size of spatial formations.}}~\cite{HV,RefYulin,RefTanese,HS,HSS,ReviewSpin} Indeed, the cubic nonlinearity associated with the Coulomb and exchange interaction between exciton polaritons (later, \textit{polaritons}), usually plays crucial role in dynamics and stability of spatial formations~\cite{World} and allows the observation of effects similar to those obtained in Kerr media such as formation of stationary and moving optical dark and bright solitons.~\cite{classic, cla} While repulsive polariton self-interaction favors the onset of dark solitons,~\cite{RefPigeon,RefSmirnov,RefPinsker} bright solitons have been produced in 
the region of dispersion corresponding to negative effective mass. 
{ {For instance, in Ref.}}~\onlinecite{Sich} { {dissipative solitons were demonstrated in an optical parametric oscillator regime. 
However, DSs in such an implementation require permanent coherent holding radiation and have low contrast due to the nonvanishing background density. Instead, in the case of DSs with incoherent (nonresonant) background radiation, proposed in Ref.}}~\onlinecite{OstrovskayaPRA86013636}{ {, inhomogeneous pumps and/or trapping potentials are required.}}

In a microcavity, polaritons can be described by a macroscopic wave function (also called the order parameter) with dynamics governed by the Gross-Pitaevskii equation (GPE).~\cite{PitaGross} In our work, we will show the contribution of the Kerr-like nonlinear terms but we will mainly focus on the DS formation due to presence of saturable absorber inside the structure exclusively. Physically, appearance of a soliton in such structures results from the increase of effective lifetime of electromagnetic field mode in the regions of high polariton density. Mathematically, saturable absorption manifests itself in the nonlinear term in the equation of motion for the order parameter, as will be shown below. 

An important feature of exciton polaritons in a microcavity is strong interaction with acoustic phonons of the crystal lattice. In the case when the DS is in the center of the nanowire, phonons may act as an additional mechanism which takes the excitations away from the central regions to the sides. Thus, phonons tend to turn a soliton-like propagation into a diffusion-like motion. This leads to a detuning and destruction of the soliton. Therefore, one has to change the settings of the pumping parameters necessary for soliton formation. 
In other words, DSs at zero temperature and at finite temperature represent two different nonlinear objects.
On the other hand, phonons may, in principle, lead to additional narrowing of the DS if the effect caused by saturable absorption is strong enough.


\section{Model}
Let us start the description of the model with the acoustic phonons--related part. 
\begin{figure}[!t]
\includegraphics[width=0.99\linewidth]{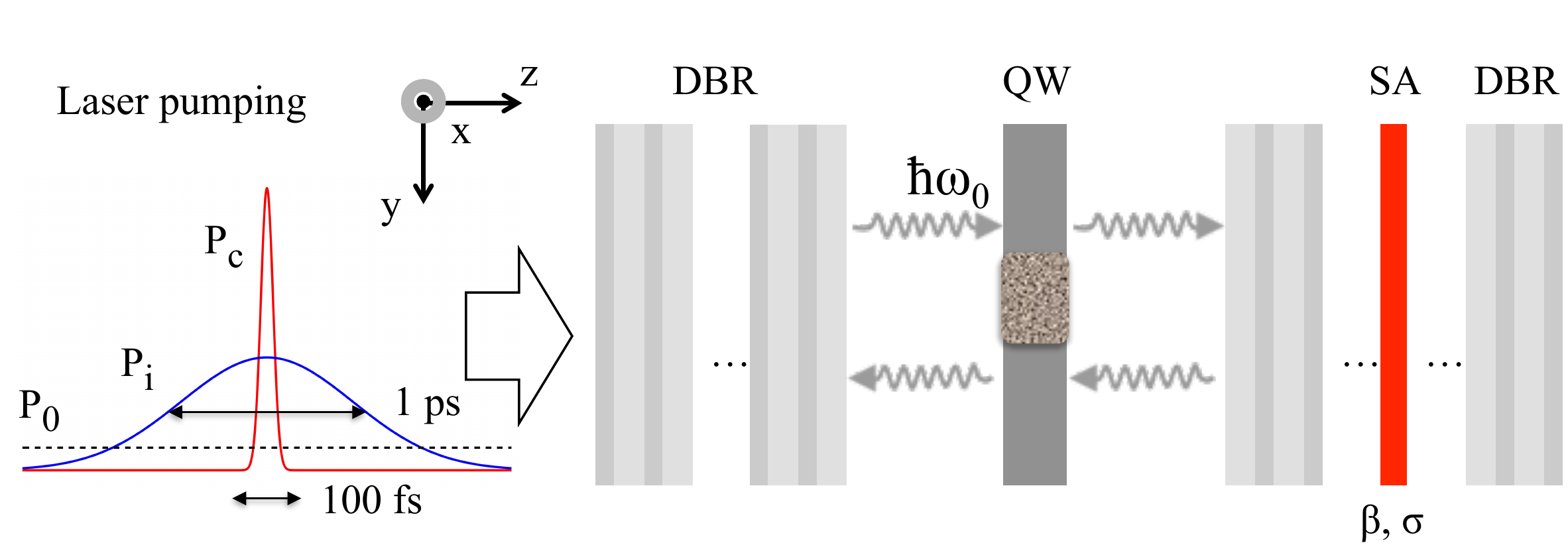}
\caption{(color online). System schematic: a single-mode semiconductor microcavity under nonresonant homogeneous excitation, $P_\textrm{0}$. The photons with frequency $\omega_\mathrm{0}$ are localized between two Bragg mirrors (DBRs) and the polaritons are localized in the quantum well (QW). Grey insert in the QW along $x$-axis represents a cross-section of a one-dimensional nanowire. The saturable absorber (SA) is embedded in one of the DBRs and described by two parameters: $\beta$ and $\sigma$. Two laser pulses: coherent, $P_\textrm{c}$, and incoherent, $P_\textrm{i}$, are used to switch the DS on and off. } 
\label{FigSystem}
\end{figure}
This part is not the core element of the model, since it is not responsible for creation of the DSs qualitatively. However, accounting for the phonons makes our simulations realistic and applicable to model experiments. We employ the theory developed in Ref.~\onlinecite{SavenkoPRL2013}. Thus, in our calculations the Fourier transform of the polariton field, $\hat\Psi(\mathbf{r},t)$, in $k$-space, $\hat a_\mathbf{k}$, is coupled to the Fourier transform of the phonon field, $\hat b_\mathbf {q}$, modeled using stochastic variables. Here $\mathbf{r}$ is a coordinate vector, $t$ is time, and $\mathbf{k}$, $\mathbf{q}$ are wave vectors of polaritons and phonons, respectively. { {Indeed, remembering that phonons represent an incoherent thermal reservoir, we can use the Markov approximation, when phonons are assumed to have a randomly varying phase.}}~\cite{Carmichael} 

We consider a system of polaritons presented in Fig.~\ref{FigSystem} and investigate one-dimensional (1D) propagation of particles along a channel inside the cavity. saturable absorber is located in one of the Bragg mirrors of the cavity, thus it represents a SESAM geometry.~\cite{RefSESAM}
%
%
The interaction with acoustic phonons comes from the Fr\"ohlich Hamiltonian,~\cite{Tassone1997}
\begin{eqnarray}
\hat{\mathcal{H}}_\textrm{int}=\sum_{\mathbf{q},k}G_{\mathbf{q}}\hat{b}_{\mathbf{q}}\hat{a}^\dagger_{k+q_x}
\hat{a}_k+G_{\mathbf{q}}^*\hat{b}^\dagger_{\mathbf{q}}\hat{a}_{k+q_x}\hat{a}^\dagger_k,
\end{eqnarray}
where $\hat{a}_k^\dagger$, $\hat{a}_k$ are polariton creation and annihilation operators in 1D. 
The phonon wave vector reads
$\mathbf{q}=\mathbf{e}_xq_x+\mathbf{e}_yq_x+\mathbf{e}_zq_z$, where $\mathbf{e}_x$,
$\mathbf{e}_y$ and $\mathbf{e}_z$ are unit vectors: $\mathbf{e}_x$ is in the 1D
wire direction, $\mathbf{e}_z$ is in the structure growth direction, $\mathbf{e}_y$ is perpendicular to both.
The phonon dispersion relation, $\hbar\omega_{\mathbf{q}}=\hbar
u\sqrt{q_x^2+q_y^2+q_z^2}$, is determined by the sound velocity,
$u$. Parameters $G_{\mathbf{q}}$ are the exciton-phonon interaction strengths evaluated elsewhere.~\cite{Hartwell2010}

The equations of motion for the polariton macroscopic wave function, $\psi$, and the reservoir occupation number, $n_\textrm{R}$, read~\cite{RefWoutersPRL991404022007, RefOstrovskayaPRL1101704072013}
\begin{eqnarray}
\label{EqGPEnR}
&&i\hbar\frac{\partial\psi(x,t)}{\partial t}={\cal F}^{-1}\left[E_k\psi_k+{\cal S}_k(t)\right]+\hbar P_\textrm{c}(x,t)\mathrm{e}^{-i\omega_\textrm{c} t}\\
\nonumber
&&~~~~~+\frac{i\hbar}{2}\left[Rn_\textrm{R}(x,t)-\gamma_0(1+\frac{\beta}{1+\sigma|\psi(x,t)|^2})\right]\psi(x,t)\\
\nonumber
&&~~~~~+\sum_k\left[{\cal T}_{-k}(t)+{\cal T}^*_k(t)\right]\psi(x,t)
+\alpha|\psi(x,t)|^2\psi(x,t);\\
&&\frac{\partial n_\textrm{R}(x,t)}{\partial t}=-(\gamma_\textrm{R}+R|\psi|^2)n_\textrm{R}+P_\textrm{0}(t)+P_\textrm{i}(x,t),
\end{eqnarray}
where ${\cal F}^{-1}$ stands for the inverse Fourier transform, $E_k$ is free dispersion, $\psi_k$ is the Fourier image of the order parameter, $P_\textrm{c}$, $P_\textrm{0}$, $P_\textrm{i}$ and $\gamma_\textrm{R}$ are the coherent pumping, incoherent reservoir homogeneous and pulsed pumping, and inverse lifetime of the reservoir, correspondingly, $R$ is the reservoir-system excitations exchange rate. The term ${\cal S}_k(t)$ corresponds to the emission of phonons by a condensate stimulated by the polariton density. 

Onwards, in the second line of Eq.~\eqref{EqGPEnR} we use the dependence of polariton inverse lifetime on their concentration, $|\psi(\mathbf{r},t)|^2$, thus the second term in the second line in square brackets is the saturable absorption-mediated term.~\cite{RefOraevskii, RefFedorov2000} { {Writing this term this way, we assume that the saturable absorber has small relaxation time and thus we can neglect its internal dynamics.}} The nonlinear dependence of inverse particle lifetime on their density stimulates the increase of the lifetime in the regions with high density and thus provides spatially-dependent lifetime-enhanced formation of localized structures. Here $\beta$ and $\sigma$ are the main parameters describing the saturable absorber, both acting on the photonic parts of polaritons: $\gamma_0(1+\beta)$ has the meaning of the effective polariton lifetime with account of the saturable absorption in the limit $|\psi|^2\rightarrow0$, $\sigma$ characterizes the saturation intensity, $\sigma\approx 1/|\psi|^2_{s}$. In the limit $\sigma|\psi|^2\gg 1$, $\gamma_c\approx\gamma_0$ and the particle lifetime is maximized, while in the opposite limit, it is minimized: $\gamma_c\approx\gamma_0(1+\beta)$.

It should be noted, given that the absorber is situated inside a DBR, both $\beta$ and $\sigma$ have direct correspondence with the photonic parts of polaritons solely (not the excitonic part). 
Indeed, this last term in the second line of Eq.~\eqref{EqGPEnR} has been adapted from the classical laser optics \cite{RefOraevskii} where a similar expression is used as a saturation term of the electric field vector in the description of a laser operation. However, due to the fact that the lifetime of polaritons is mostly determined by the lifetime of photons in the cavity (inverse lifetimes of photons and excitons scales as $\gamma_{0}\approx 10\gamma_{X}$) and here this condition is additionally strengthened (due to additional photonic losses caused by the saturable absorber), we conclude that lifetime of polaritons is also determined by the photonic lifetime and therefore formula~\eqref{EqGPEnR} is feasible. 

The stochastic terms, ${\cal T}_{q_x}$, in the last line of Eq.~\eqref{EqGPEnR} are defined by the correlations~\cite{SavenkoPRL2013}
\begin{align}
\left<{\cal T}_{q_x}^*(t){\cal T}_{q_x^\prime}(t^\prime)\right>&=\sum_{q_y,q_y}\left|G_{{q_x,q_y,q_z}}\right|^2N_{q_x,q_y,q_z}\delta_{q_x,q_x^\prime}\delta(t-t^\prime);\notag\\
\left<{\cal T}_{q_x}(t){\cal T}_{q_x^\prime}(t^\prime)\right>&=
\left<{\cal T}_{q_x}^*(t){\cal T}_{q_x^\prime}^*(t^\prime)\right>=0,
\label{EqThermal}
\end{align}
where $N_{\mathbf{q}}$ is the number of phonons in the state with a wave vector $\mathbf{q}$ determined by the temperature of the system. The proportionality of the thermal part to the first power of $\psi$ in Eq.~\eqref{EqGPEnR} leads to spontaneous scattering processes. Averaging over $q_x$ and $q_y$, we obtain 1D polariton dynamics along the wire. 

\begin{figure}[!t]
\includegraphics[width=1.0\linewidth]{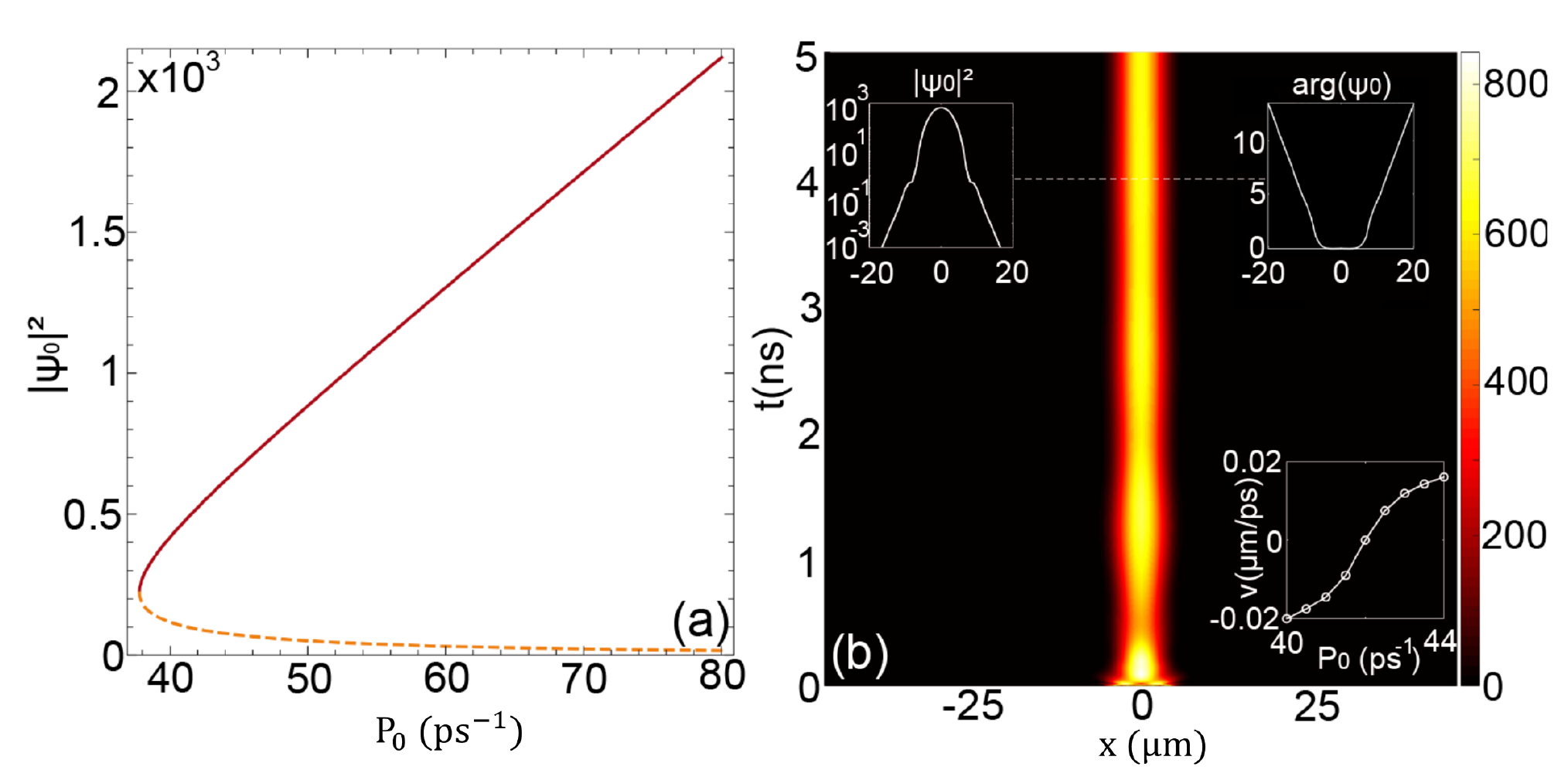}
\caption{(color online). (a) Bistability in the steady state: dependence of the exciton-polariton concentration, $|\psi_0|^2$, on the intensity of nonresonant pump, $P_0$, for the parameters: $\gamma_0=0.025$ $ps^{-1}$, $R=5\cdot 10^{-6}$ $ps^{-1}$, $\gamma_\textrm{R}=1/200$ $ps^{-1}$, $\beta=5$, $\sigma=0.1/dx/w_y$, $dx$ is discretization length, $w_y$ is the 1D microwire width. The $0$-branch corresponding to stable trivial solution is not shown.
(b) DS formation in the absence of the particle self-scattering, $\alpha=0$; the intensity of pump is $P_0=42$ ps$^{-1}$. The colormap shows the density in $\mu$m$^{-2}$.
Upper left-most inset shows the DS density profile (with the width $\approx$ 10 $\mu$m) in the steady state. 
Upper right-most inset illustrates the phase distribution. 
Lower inset illustrates the switching wave velocity dependence on the intensity of pump. At some value of pump, the switching waves stop ($v=0$).} 
\label{FigBistabiitySoliton}
\end{figure}

In the case of homogeneous excitation of the system [$P_0=\textrm{Const}(x)$ and thus $n_\textrm{R}=n_\textrm{R0}=\textrm{Const}(x)$] in a steady state (when $n_\textrm{R}=\textrm{Const}(t)$, $\psi(x,t)=\psi_0(x)exp(-i\omega t)$ and thus $|\psi(x,t)|^2=|\psi_0|^2$), splitting real and imaginary parts in \eqref{EqGPEnR} we obtain the following equations:
\begin{eqnarray}
\label{EqChemicalPotential}
\hbar\omega&=&\alpha|\psi_0|^2;\\
\label{EqSteadyPsi}
0&=&\left[Rn_\textrm{R0}-\gamma_0\left(1+\frac{\beta}{1+\sigma|\psi_0|^2}\right)
\right]\psi_0;\\
\label{EqSteadynR}
0&=&-(\gamma_\textrm{R}+R|\psi_0|^2)n_\textrm{R0}+P_0.
\end{eqnarray}
Eq.~\eqref{EqChemicalPotential} fixes the chemical potential of the particles, $\mu=\alpha|\psi_0|^2$, 
Eq.~\eqref{EqSteadyPsi} is the gain condition which defines the homogeneous polariton occupation directly linked to the reservoir particle number in a steady state by Eq.~\eqref{EqSteadynR}. Apart from the trivial (no-lasing) solution $(|\psi_0|^2)_{0}=0$, we find $(|\psi_0|^2)_{1,2}=(-B\pm\sqrt{B^2-4AC})/(2A)$,
%
%
where
$A=\gamma_0\sigma R$, $B=-(RP_0\sigma-\gamma_0\gamma_\textrm{R}\sigma-\gamma_0R-\gamma_0\beta R)$ and $C=-(RP_0-\gamma_0\gamma_\textrm{R}-\gamma_0\gamma_\textrm{R}\beta)$. These three solutions simultaneously exist in the region $P_\textrm{min}< P_0 < P_\textrm{max}$, however, only two of them are stable: $(|\psi_0|^2)_{0,1}$ forming the bistability, see Fig.~\ref{FigBistabiitySoliton}(a). 
A linear stability analysis of the solutions is presented in Appendix A.


\section{Dissipative soliton formation}
The possibility of changeover between two stable solutions in some range of pumps leads to formation of a \textit{switching wave} that essentially occurs at the boundary between the regions with different particle number. The switching wave favours one of the solutions and tends to establish homogeneity of the system. 
The velocity of propagation of the switching wave, $v$, as a function of $P_\textrm{0}$ vanishes at the so-called Maxwell's value of pump, $P_\textrm{M}$ [see Fig.~\ref{FigBistabiitySoliton}(b)]. In the vicinity of this value, two spatial domains coexist and a spatial soliton can be formed.~\cite{RefRosFed1992,RefRosKho1988}
Here we neglect the phonon-related interaction. The tradeoff between (i) gain and free dispersion terms which together favour diffusion-like spreading of particles and (ii) nonlinear losses favouring localisation may result in a bright DS. It occurs in the range of pump intensities, $P_\textrm{min}<P_\textrm{M}<P_\textrm{max}$, at which none of the processes (i), (ii) surpasses the other.
Instead, at $P<P_\textrm{min}$ the system collapses towards the no-lasing solution, whereas at $P>P_\textrm{max}$ the final state represents homogeneous profile with high particle concentration.

\begin{figure}[t!]
\includegraphics[width=0.99\linewidth]{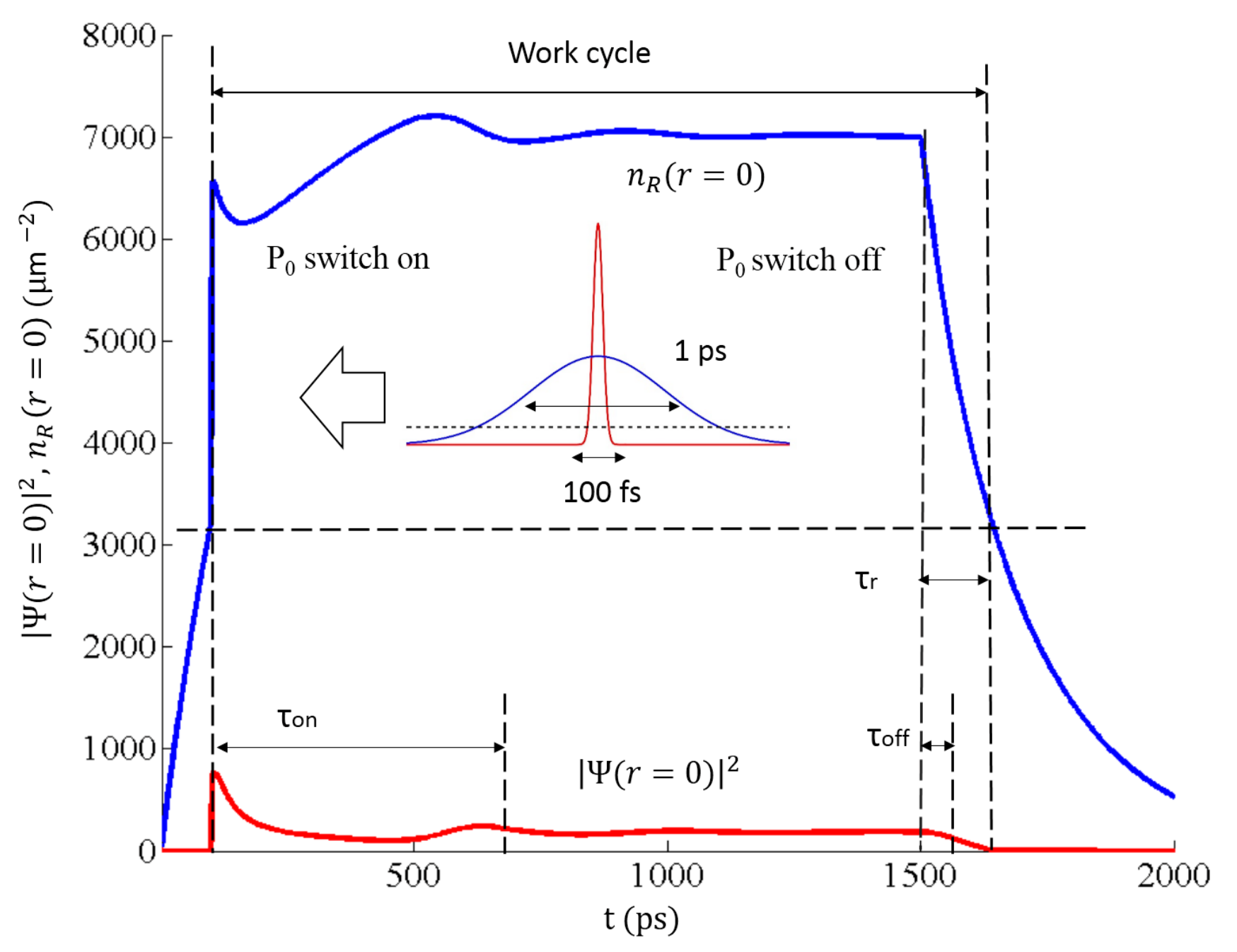} 
\caption{(color online). DS protocol for arbitrary state at $T=0$: $|\psi|^2$ and $n_\textrm{R}$ at $x=0$ as a functions of $t$. {{Inset shows the parameters of the pulses. Here $t_\textrm{c}=100$ fs, $t_\textrm{i}=1$ ps, and their spatial widths $r_\textrm{c}=10$ $\mu$m, $r_\textrm{i}=50$ $\mu$m; $P_i=9301$, $P_c=1066$ ps$^{-1}$.}} The switch on time is $\tau_\textrm{on}=$600 ps, switch off time is $\tau_\textrm{off}$=94 ps. At $t\approx 1700$ ps (when $n_\textrm{R}(0,t)\approx 3100$ $\mu$m $^{-2}$, $|\psi|^2=0$), the system can be pulsed again, and the protocol returns to its starting point, $t=100$ ps. The ``return'' time is $\tau_\textrm{r}=200$ ps, not shown in figure.} 
\label{FigProtocol1}
\end{figure}

Time of creation and lifetime of the DS critically depend on the pumping parameters. Let us consider two different protocols of DS creation at zero temperature.
Initially, we  create a Gaussian density profile in the centre of the sample by a short strong laser pulse $P_c(r,t)=P_c \mathrm{e}^{{-t^2}/{t_c^2}}\mathrm{e}^{{-r^2}/{r_c^2}} $, thus preparing two spatial regions with different particle concentration (see Fig.~\ref{FigProtocol1}). The second pulse which also has a Gaussian density profile is used for the nonresonant excitation, $P_\textrm{i}(r,t)=P_\textrm{0}+P^{(0)}_\textrm{i} \mathrm{e}^{{-t^2}/{t_i^2}}\mathrm{e}^{{-r^2}/{r_i^2}}$. Further, the system evolves under background homogeneous nonresonant excitation, $P_0$, solely. 

In the protocol (see Fig.~\ref{FigProtocol1}) we see that two short pulses issued at the same time (centered at 100 ps) lead to critical changes in the balance between the system and the reservoir. 
Coherent pulse in the center of the wire leads to an increase of the number of polaritons, $|\psi|^2$, and more efficient exchange with the reservoir, $n_\textrm{R}$. 
During first 100 ps of the protocol, $|\psi(x,t)|^2$ remains zero. Thus, in Fig.~\ref{FigProtocol1} we observe a quasi-linear dependence of $n_\textrm{R}$ on $t$ under homogeneous pumping $P_\textrm{0}$.

\begin{figure}[t!]
\includegraphics[width=0.99\linewidth]{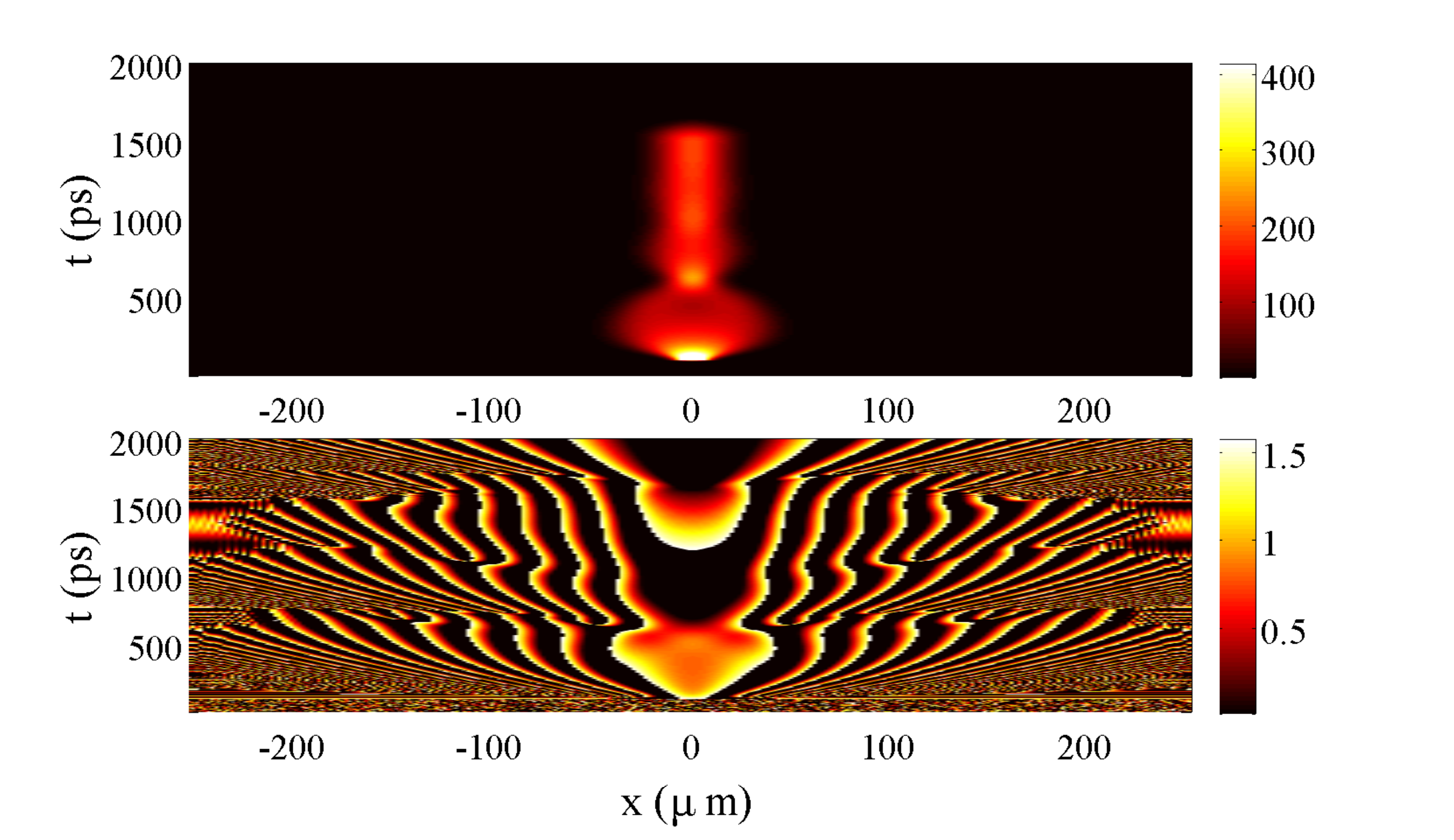} 
\caption{(color online). Dissipative soliton creation and destruction. 
Upper panel: $|\psi(x,t)|^2$ in $\mu$m$^{-2}$ as a function of coordinate along the nanowire, $x$, and time, $t$ at $T=0$ K. 
Lower panel: the phase of the condensate particles as a function of $x$ and $t$. 
The intensity of the background pump is $P_\textrm{0}=42$ ps$^{-1}$. 
Coherent pumping is switched on at $t=100$ ps (see also Fig. 1 and 3 for the details of the protocol). At $100<t<500$ ps, two spatial domains coexist. Standing soliton is formed after $\approx$ 750 ps and it is stable during time until we switch it off at 1500 ps. } 
\label{FigSwitchOnOff}
\end{figure}

The soliton can be destroyed if we switch off the constant background pumping, $P_\textrm{0}$. In this case, reduction of the number of particles in the reservoir leads to reduction of particles number in the system and the DS disappears fast, see Fig.~\ref{FigSwitchOnOff} and Fig.~\ref{FigProtocol1} after 1500 ps. The period of time in which $|\psi(0,t)|^2$ decreases exponentially, is called the time of switch off, $\tau_\textrm{off}$. In our calculation we found that $\tau_\textrm{off}= 94$ ps. Interesting to note that it is possible to achieve $\tau_\textrm{on}<\tau_\textrm{off}$. After $\tau_\textrm{r}=200$ ps, the protocol returns to its starting point, $t=100$ ps.
Thus, varying amplitude, width and duration of the pumping sources we switch from a linear wave to a soliton-like motion, see Fig.~\ref{FigSwitchOnOff}.
Selection of pumping parameters allows us to achieve steady DS on a ns-scale time. Conventionally, we denote the time of switch on, $\tau_\textrm{on}$, as a time between the center of the initial pump pulse and the moment when the density of polaritons stops to fluctuate. When $|\psi|^2$ and $n_\textrm{R}$ stabilize, the DS forms.

\begin{figure}[b!]
\includegraphics[width=1.0\linewidth]{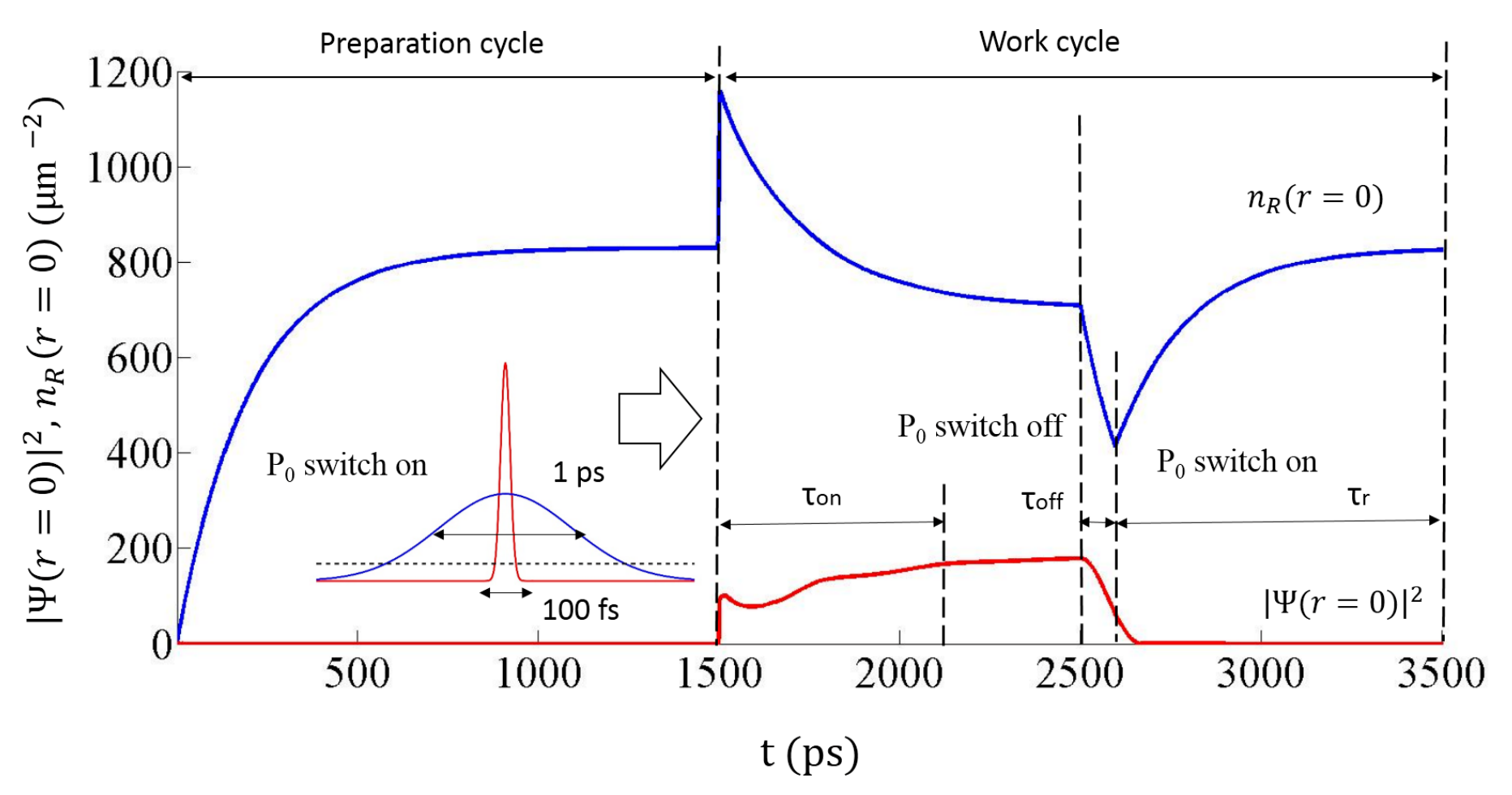} 
\caption{(color online). DS protocol for saturation reservoir state at $T=0$: $|\psi|^2$ and $n_\textrm{R}$ at $x=0$ as a functions of $t$. At $1500$ ps (when $n_\textrm{R}(0,t)\approx 800$ $\mu$m $^{-2}$, $|\psi|^2=0$) reservoir comes to homogeneous saturated state, the system can be pulsed. Inset shows the parameters of the pulses. Here $t_\textrm{c}=100$ fs, $t_\textrm{i}=1$ ps, and their spatial widths are $r_\textrm{c}=10$ $\mu$m, $r_\textrm{i}=50$ $\mu$m; $P_i=1890$, $P_c=365$ ps$^{-1}$. The switch on time is $\tau_\textrm{on}=$600 ps. At $2500$ ps constant pumping is switch off (switch off time is $\tau_\textrm{off}\approx 100 ps$) and switched on after reservoir population decreases exponentially. The return time is $\tau_\textrm{r}=900$ ps.} 
\label{FigProtocol2}
\end{figure}

In the protocol (see Fig.~\ref{FigProtocol2}) we see that during first 1500 ps of the protocol, $|\psi(x,t)|^2$ remains zero and $n_\textrm{R}$ reaches the saturation. Then, two short pulses are issued at 1500 ps.
Later, the DS is formed, which can be destroyed if we switch off the constant background pumping, $P_\textrm{0}$. After about $\tau_\textrm{r}=700$ ps, the protocol returns to its starting point.

The second protocol is more useful from the applications viewpoint. We keep the reservoir at high-number state and it takes less time for the DS to be established. Moreover, the intensities of pumps required to ignite the second protocol are much smaller, $P_i=1890$, $P_c=365$ vs $P_i=9301$, $P_c=1066$ ps$^{-1}$ [compare captions of Fig.~\ref{FigProtocol2} and Fig.~\ref{FigProtocol1}].


\section{Influence of scattering on acoustic phonons}
\begin{figure}[t!]
\includegraphics[width=0.99\linewidth]{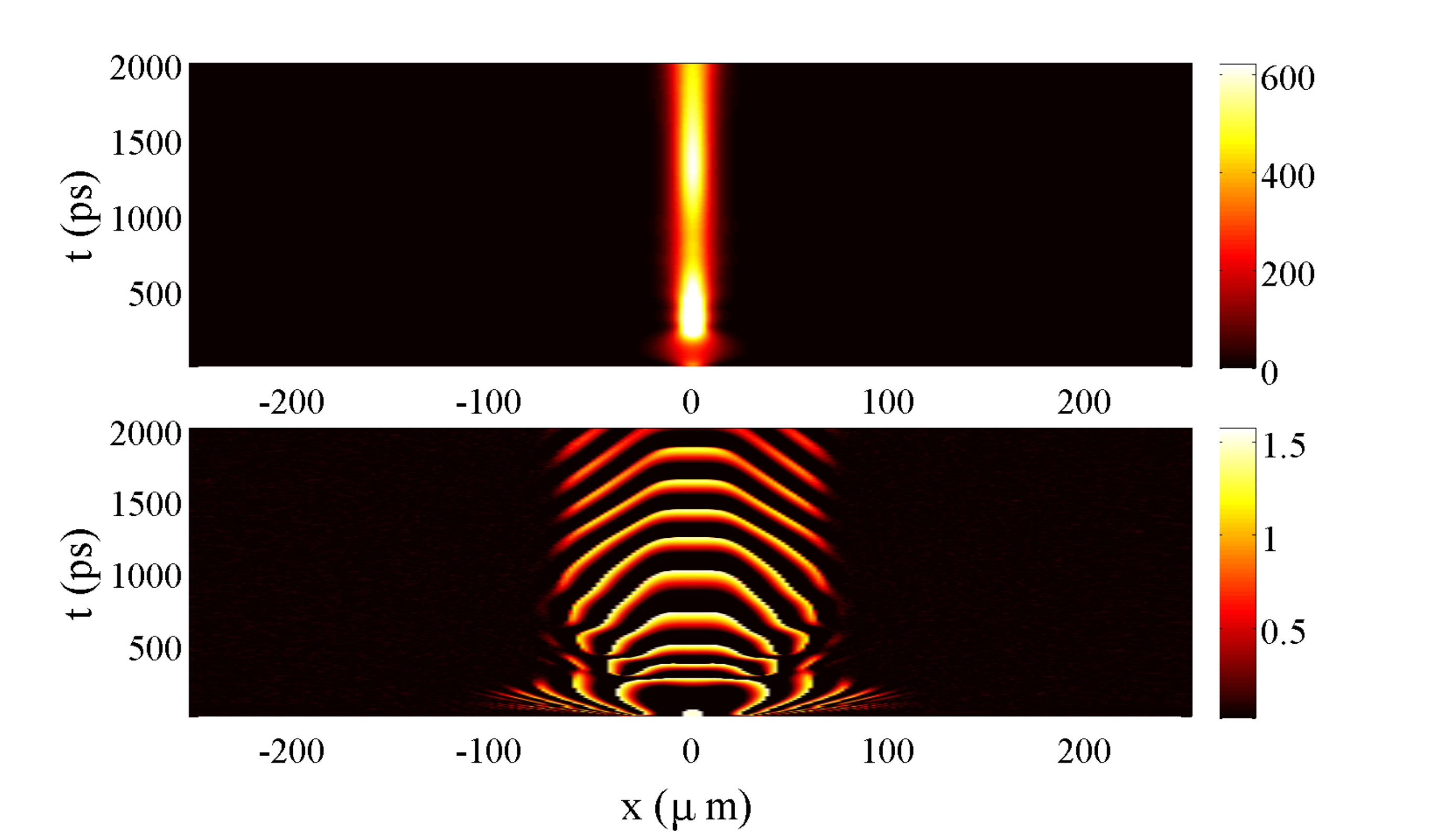} 
\caption{(color online). Illustration of the temperature influence on the DS behavior. Upper panel: $|\psi(x,t)|^2$ in $\mu$m$^{-2}$ as a function of coordinate along the nanowire, $x$, and time, $t$ at temperature $T=15$ K. 
Lower panel: the phase of the condensate particles as a function of $x$ and $t$. 
The intensity of the background pump is $P_\textrm{0}=43$ ps$^{-1}$.}
\label{FigFiniteTemperature}
\end{figure}

Let us further account for the finite temperature ($T>0$). The results of modeling for $T=15$ K are presented in Fig.~\ref{FigFiniteTemperature}. The relaxation of energy of polaritons (thermalisation) caused by the interaction with the bath of phonons leads to a narrowing of the DS profile.  
This happens since exciton--phonon interaction leads to more intensive hauling of particles from the center of the QW similar to diffusion in classical systems. However, in the side regions the lifetime of polaritons is lower than in the center ($x=0$), hence the spatial extent of the DS is smaller.
The DS is now a (fourfold) trade off between the (i) gain, dispersion, and particle--phonon interaction and (ii) nonlinear losses. At moderate temperatures, phonon-mediated relaxation does not break the DS: the narrowed DS is still quite stable, meaning that (ii) can compensate processes (i). 

\begin{figure}[t!]
\includegraphics[width=0.99\linewidth]{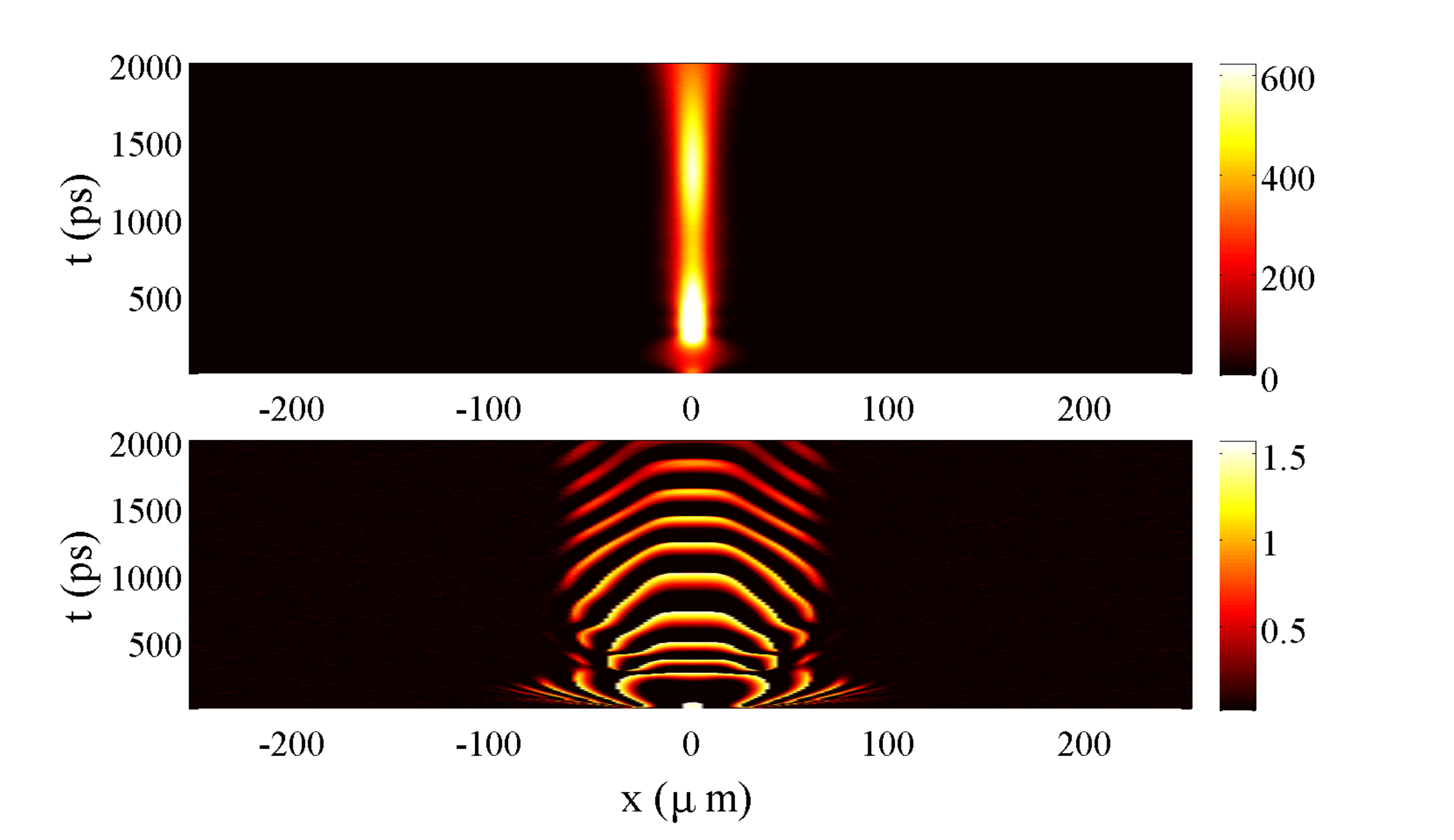} 
\caption{(color online). Illustration of decay of the DS due to interaction with acoustic phonons. Upper panel: $|\psi(x,t)|^2$ in $\mu$m$^{-2}$ as a function of coordinate along the nanowire, $x$, and time, $t$ at temperature $T=25$ K. $P_\textrm{0}=43$ ps$^{-1}$.
Lower panel: the phase of the condensate particles as a function of $x$ and $t$.}
\label{FigFiniteTemperature2}
\end{figure}

Further we increase the temperature, $T=25$ K, see Fig.~\ref{FigFiniteTemperature2}, where we used the same parameters as in Fig.~\ref{FigFiniteTemperature} apart from the temperature. 
We observe the soliton collapse: after about 1600 ps the particle density in the center of the 1D channel is reduced, and the density of energy (the phase presented in the lower panel of Fig.~\ref{FigFiniteTemperature2}) is also decreases. The phase of the wave function undergoes spatial redistribution. 

Important to note, the phonons effectively play a role of a noise in the system. Usually, noise leads to stochastic shift of the soliton trajectory resulting in oscillations of the particle number maxima.~\cite{Nesterov2015} However, it follows from our simulations that this noise does not lead to shift or oscillations of the trajectory of the DS, it keeps still, even at $T=25$ K.

It should also be noted, that in the case of nonresonant (incoherent) excitation of the system, which is the case in our setup, polaritons initially accumulate at high-energy state close to the excitonic part of the polariton dispersion. From those energies, they scatter towards the lowest-energy state via phonon-mediated relaxation and self-scattering and then can form the condensate. 
Therefore, strictly speaking, without phonon-mediated energy scattering, polaritons may never form a BEC, required for operation of our setup. 
{{Hence, phonons have positive influence to the formation of the DS: at moderate temperatures, phonons decrease the time of the condensation.
It should additionally be stressed, that phonons result in lower switch-on time, $\tau_{on}$, of the DS since they effectively decrease the lifetime of polaritons in the side regions of the sample. And since the switching wave effectively moves faster, the system reaches the steady state sooner.}}

While in the BEC state, polaritons are still influenced by the phonons which usually cause broadening of the linewidth of the spectrum at finite temperatures, mediating transitions from the ground state (the lowest-energy state) to the nearest low-energy states. This process leads to general decrease of coherence of the system at non-zero temperatures and tends to destroy the DS.


\section{Influence of polariton self-scattering}
\begin{figure}[!t]
\includegraphics[width=1.0\linewidth]{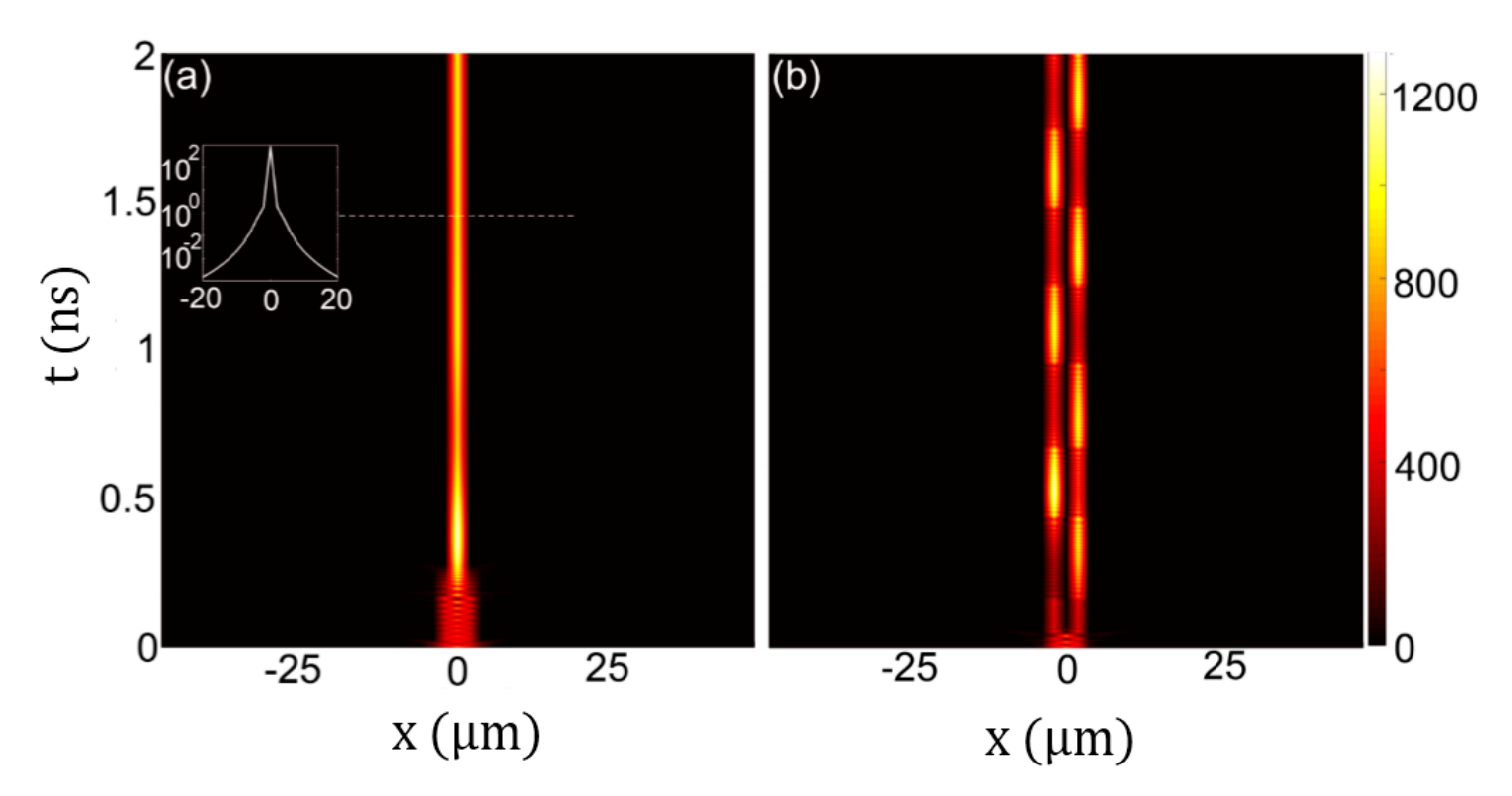}
\caption{(color online). DS formation with account of the particles self-scattering, $\alpha=$15 $\mu$eV$/\mu$m$^2$. Parameters of the calculation are the same as in Fig.~\ref{FigBistabiitySoliton}, the pump intensity is $P_0=48$ ps$^{-1}$. 
(a) Single soliton formation for 1 $\mu$m initial Gaussian wavepacket; (b) creation of pair of solitons for 2 $\mu$m initial wavepacket.
Inset in (a) demonstrates logarithmic DS profile (with the width less than 1 $\mu$m) in the steady state. 
See main text for details.} 
\label{FigSolitonInteraction}
\end{figure}

Accounting for the particle-particle scattering drastically changes the properties of DSs, 
see Fig.~\ref{FigSolitonInteraction}. In this section, we will neglect the contribution of scattering on acoustic phonons to demonstrate the effect of self-scattering exclusively.
We use $\alpha=0.5\cdot 6\cdot E_ba_B^2/dx/w_y$, where $E_\textrm{b}$ is the exciton binding energy, $a_\textrm{B}$ is its Bohr radius, $dx$, $w_y$ are the 1D discretization element length and the width of the 1D microwire (0.5 and 2 $\mu$m, correspondingly). 
We observe a remarkable alteration of the DS properties [compare Fig.~\ref{FigSolitonInteraction}(a) and Fig.~\ref{FigBistabiitySoliton}(b)]. 
The DS is now a (fourfold) trade off between the (i) gain, dispersion, and particle repulsion (since $\alpha>0$) and (ii) nonlinear losses, where processes (i) try to stretch and wash the soliton out. Obviously, $P_\textrm{M}$ is displaced in that case (48 instead of 42 ps$^{-1}$). 

It should be emphasized that the self-interaction turns out beneficial as it promotes additional constriction of the spatial profile and the soliton width is now in the sub-micron range. This reduction of the spatial formation is the direct result of the ancillary particle-particle scattering. Indeed, it imposes a smaller (comparing to the $\alpha=0$ case) peak extent for the balance condition to occur. 

Onwards in Fig.~\ref{FigSolitonInteraction}(b) we demonstrate that variation of the initial ($t=0$) profile allows one to create various spatial patterns in the steady state ranging from a single peak [panel (a)] to multipeaks [panel (b)]. Noteworthy is the population exchange between the solitons clearly observable in Fig. \ref{FigSolitonInteraction}(b), similar to Josephson-like oscillations. We attribute this effect to the soliton tails interaction that produce a potential barrier between the solitons. Moreover, self-scattering results in decrease of the switch-on time of the soliton, and here it is $\tau_\textrm{on}=18$ ps (compare with 100-s ps, as in Fig.~\ref{FigProtocol1}).


%
%
%
%
%
%



\section{Conclusions}
In summary, we considered a system of exciton polaritons in contact with a bath of acoustic phonons in a semiconductor microcavity with an embedded saturable absorber. We showed that even at finite temperature it is possible to observe the nonlinear behavior of the system with the emergence of a dissipative soliton. Moreover, the soliton exists if we account for nonlinear polariton-polariton self-scattering. Further, we  proposed protocols which allow to create dissipative solitons using two short laser pulses: one resonant and another one incoherent. We demonstrate that interaction with phonons and self-interaction lead to decrease of the switch-on lifetime of the solitons and decrease of the spatial extent. We believe that these calculations can be used in future experiments on dissipative solitons and creation of controlled bistability at nonresonant excitation. 

\section*{Acknowledgements}
We thank S. Fedorov, T. Liew and R. Lake for useful discussions. We acknowledge financial support from the Academy of Finland through its Centre of Excellence Programs (Projects No. 250280 and No. 251748); the Government of Russian Federation, grant 074-U01; and the Dynasty Foundation. 
The calculations presented here were partly performed using supercomputer facilities within the
Aalto University School of Science “Science-IT” project.



\section*{Appendix A. Linear stability analysis}
Here we will omit the phonon-related processes. In the steady state (when $n_\textrm{R}=\textrm{\textrm{Const}}(t)$, $\psi(\mathbf{r},t)=\psi_0(\mathbf{r})exp(-i\omega t)$ and thus $|\psi(\mathbf{r},t)|^2=|\psi_0|^2$), assuming homogeneous excitation of the system [$P(\mathbf{r})=P_0$ and thus $n_\textrm{R}=n_\textrm{R0}=\textrm{\textrm{Const}}(\mathbf{r})$], we obtain the following equaitons:
\begin{eqnarray}
\label{EqGPEnRss}
\hbar\omega\psi_0&=&\alpha|\psi_0|^2\psi_0+\frac{i\hbar}{2}(Rn_\textrm{R0}-\gamma_c(|\psi_0|^2))\psi_0;\\
\nonumber
0&=&-(\gamma_\textrm{R}+R|\psi_0|^2)n_\textrm{R0}+P_0.
\end{eqnarray}
Or splitting the real and imaginary parts in Eq.~\eqref{EqGPEnRss}, we find
\begin{eqnarray}
\label{EqGPEnRssSplit}
\hbar\omega&=&\alpha|\psi_0|^2;\\
\nonumber
0&=&\left[Rn_\textrm{R0}-\gamma_c(|\psi_0|^2)\right]\psi_0;\\
\nonumber
0&=&-(\gamma_\textrm{R}+R|\psi_0|^2)n_\textrm{R0}+P_0.
\end{eqnarray}
From equations above, we find that
\begin{eqnarray}
\label{EqPsi2}
(|\psi_0|^2)_{1,2}=\frac{-B\pm\sqrt{B^2-4AC}}{2A},
\end{eqnarray}
where
$A=\gamma_0\sigma R$, $B=-(RP_0\sigma-\gamma_0\gamma_\textrm{R}\sigma-\gamma_0R-\gamma_0\beta R)$, $C=-(RP_0-\gamma_0\gamma_\textrm{R}-\gamma_0\gamma_\textrm{R}\beta)$. Thus, two solutions are possible.
The bistability is presented in Fig.~1 of the main text. Oone should be careful with the choice of the pump intensity (that comes from the quadratic equation). For example, the condition $B^2-4AC>0$ should be satisfied and the following inequality be valid:
\begin{eqnarray}
\label{PMaxMin}
\frac{\gamma_0(\sigma\gamma_\textrm{R}+R+\beta R)}{\sigma R}<P_0<\frac{\gamma_0\gamma_\textrm{R}\sigma(1+\beta)}{\sigma R}.
\end{eqnarray}
The first solution of Eq. (9) is trivial (the so-called generation-free or non-lasing solution): $\psi=0$. Further, in \eqref{EqGPEnR} let us assume $|\psi|^2\rightarrow 0$.
Then we can easily neglect the nonlinear terms in the equation \eqref{EqGPEnR} and for $m=|\psi|$ we have (the free propagation term vanishes after taking the absolute value):
\begin{eqnarray}
\nonumber
\frac{dm}{dt}=\frac{1}{2}\left(\frac{P_0R}{\gamma_\textrm{R}}-\gamma_0(1+\beta)\right)m,
\end{eqnarray}
where $m$ obviously vanishes with $t$ if
\begin{eqnarray}
\label{EqNoGeneration}
P_0R-\gamma_\textrm{R}\gamma_0(1+\beta)<0.
\end{eqnarray}
This is the stability condition for the no-lasing regime.

\begin{widetext}

To investigate the regimes of generation, let us introduce the disturbed variables
\begin{eqnarray}
\label{EqPsinRstability}
\psi&=&\psi_0\mathrm{e}^{-i\omega t}(1+a\mathrm{e}^++b^*\mathrm{e}^-);\\
\nonumber
n_\textrm{R}&=&n_\textrm{R0}(1+(\delta n)\mathrm{e}^++(\delta n^*)\mathrm{e}^-),
\end{eqnarray}
where
\begin{eqnarray}
\nonumber
&&\mathrm{e}^+=\mathrm{e}^{i\mathbf{k}_{\perp}\mathbf{r}_{\perp}+\gamma t}; \mathrm{e}^-=(\mathrm{e}^+)^*;\\
\nonumber
&&a,b,\delta n\rightarrow 0
\end{eqnarray}
are treated as independent small perturbations. Then, the time derivative reads (linearizing over the small variables everywhere below)
\begin{eqnarray}
\label{EqTimeDerivativeStability}
i\hbar\frac{\partial\psi}{\partial t}=\hbar\omega\psi_0\mathrm{e}^{-i\omega t}
\left[1+(1+i\frac{\gamma}{\omega})a\mathrm{e}^++(1+i\frac{\gamma^*}{\omega})b^*\mathrm{e}^-\right].
\end{eqnarray}
The kinetic energy term is
\begin{eqnarray}
\label{EqKineticEnergyStability}
\frac{\hbar^2}{2m}\Delta_{\perp}\psi=\frac{\hbar^2}{2m}\psi_0\mathrm{e}^{-i\omega t}
(-k_{\perp}^2)
\left[(a\mathrm{e}^++b^*\mathrm{e}^-\right].
\end{eqnarray}
We can calculate that the first nonlinear term in Eq.~(1) reads
\begin{eqnarray}
\label{EqNonlinearTermStability}
\alpha|\psi|^2\psi=\alpha|\psi_0|^2\psi_0\mathrm{e}^{-i\omega t}
\left[1+(2a+b)\mathrm{e}^++(a^*+2b^*)\mathrm{e}^-\right].
\end{eqnarray}
The polaritons-reservoir interaction term read
\begin{eqnarray}
\label{EqReservoirTermStability}
\frac{i\hbar}{2}Rn_\textrm{R}\psi=\frac{i\hbar}{2}Rn_\textrm{R0}\psi_0\mathrm{e}^{-i\omega t}
\left[1+(a+\delta n)\mathrm{e}^++(b^*+\delta n^*)\mathrm{e}^-\right].
\end{eqnarray}
Lifetime-dependent terms read
\begin{eqnarray}
\label{EqLifetimeTermStability}
\frac{i\hbar}{2}\gamma_0\psi&=&\frac{i\hbar}{2}\gamma_0\psi_0\mathrm{e}^{-i\omega t}
\left[1+a\mathrm{e}^++b^*\mathrm{e}^-\right];\\
\nonumber
\frac{i\hbar}{2}\gamma_0\frac{\beta}{1+\sigma|\psi|^2}\psi&=&
\frac{i\hbar}{2}\gamma_0\frac{\beta}{1+\sigma|\psi_0|^2}\psi_0\mathrm{e}^{-i\omega t}
\left[1+\{a-\frac{\sigma|\psi_0|^2}{1+\sigma|\psi_0|^2}(a+b)\}\mathrm{e}^++\{b^*-\frac{\sigma|\psi_0|^2}{1+\sigma|\psi_0|^2}(a^*+b^*)\}\mathrm{e}^-\right].
\end{eqnarray}
In the equation for the reservoir,
\begin{eqnarray}
\label{EqReservTermStability}
\gamma_\textrm{R}n_\textrm{R}&=&\gamma_\textrm{R}n_\textrm{R0}[1+\delta n\mathrm{e}^++\delta n^*\mathrm{e}^-];\\
\nonumber
R|\psi|^2n_\textrm{R}&=&R|\psi_0|^2n_\textrm{R0}[1+(a+b+\delta n)\mathrm{e}^++(a^*+b^*+\delta n^*)\mathrm{e}^-].
\end{eqnarray}
Substituting all these terms in Eq.~\eqref{EqGPEnR},
we obtain (after minor reductions) the following system of equations on $a$, $b^*$, $\delta n$ and conjugates:
\begin{eqnarray}
\label{EqGPEnRStability}
0&=&\hbar\omega[(1+i\frac{\gamma}{\omega})a\mathrm{e}^++(1+i\frac{\gamma^*}{\omega})b^*\mathrm{e}^-]
-\frac{\hbar^2}{2m}[k_{\perp}^2(a\mathrm{e}^++b^*\mathrm{e}^-)]-\alpha|\psi_0|^2[(2a+b)\mathrm{e}^++(a^*+2b^*)\mathrm{e}^-]-\\
\nonumber
&&-\frac{i\hbar}{2}Rn_\textrm{R0}[(a+\delta n)\mathrm{e}^++(b^*+\delta n^*)\mathrm{e}^-]+\\
\nonumber
&&
+\frac{i\hbar}{2}\gamma_0[a\mathrm{e}^++b^*\mathrm{e}^-]
+\frac{i\hbar}{2}\gamma_0\frac{\beta}{1+\sigma|\psi|^2}
[
\{a-\frac{\sigma|\psi_0|^2}{1+\sigma|\psi_0|^2}(a+b)\}\mathrm{e}^++\{b^*-\frac{\sigma|\psi_0|^2}{1+\sigma|\psi_0|^2}(a^*+b^*)\}\mathrm{e}^-
];\\
\nonumber
0&=&[\gamma\delta n\mathrm{e}^++\gamma^*\delta n^*\mathrm{e}^-]+\gamma_\textrm{R}[\delta n \mathrm{e}^++\delta n^*\mathrm{e}^-]+R|\psi_0|^2[(a+b+\delta n)\mathrm{e}^++(a^*+b^*+\delta n^*)\mathrm{e}^-].
\end{eqnarray}
Splitting here the $\mathrm{e}^+$ and $\mathrm{e}^-$ terms, we obtain:
\begin{eqnarray}
\nonumber
0&=&\hbar\omega[(1+i\frac{\gamma}{\omega})a]
-\frac{\hbar^2}{2m}[k_{\perp}^2a]-\alpha|\psi_0|^2[2a+b]-\frac{i\hbar}{2}Rn_\textrm{R0}[a+\delta n]
+\frac{i\hbar}{2}\gamma_0[a]
+\frac{i\hbar}{2}\gamma_0\frac{\beta}{1+\sigma|\psi|^2}
[a-\frac{\sigma|\psi_0|^2}{1+\sigma|\psi_0|^2}(a+b)];\\
\nonumber
0&=&\hbar\omega[(1+i\frac{\gamma^*}{\omega})b^*]
-\frac{\hbar^2}{2m}[k_{\perp}^2b^*]-\alpha|\psi_0|^2[a^*+2b^*]-\frac{i\hbar}{2}Rn_\textrm{R0}[b^*+\delta n^*]
+\frac{i\hbar}{2}\gamma_0[b^*]\\
\nonumber
&&+\frac{i\hbar}{2}\gamma_0\frac{\beta}{1+\sigma|\psi_0|^2}
[
b^*-\frac{\sigma|\psi_0|^2}{1+\sigma|\psi_0|^2}(a^*+b^*)
];\\
\nonumber
0&=&[\gamma\delta n]+\gamma_\textrm{R}[\delta n]+R|\psi_0|^2[(a+b+\delta n)];\\
\nonumber
0&=&[\gamma^*\delta n^*]+\gamma_\textrm{R}[\delta n^*]+R|\psi_0|^2[(a^*+b^*+\delta n^*)].
\end{eqnarray}
Taking the complex conjugate of the second equation in the system above and realizing a similarity between the last two equations above
(they are just complex conjugates of each other), we finally come up with a closed system of three equations for $a$, $b$, and $\delta n$:
\begin{eqnarray}
\label{EqStabilitySystem}
0&=&\left[\hbar\omega(1+i\frac{\gamma}{\omega})
-\frac{\hbar^2}{2m}k_{\perp}^2-2\alpha|\psi_0|^2-\frac{i\hbar}{2}Rn_\textrm{R0}
+\frac{i\hbar}{2}\gamma_0(1+\frac{\beta}{1+\sigma|\psi_0|^2}-\frac{\beta\sigma|\psi_0|^2}{(1+\sigma|\psi_0|^2)^2})\right]a \\
\nonumber
&&+ \left[-\alpha|\psi_0|^2-\frac{i\hbar}{2}\gamma_0\frac{\beta\sigma|\psi_0|^2}{(1+\sigma|\psi_0|^2)^2}\right]b
+\left[-\frac{i\hbar}{2}Rn_\textrm{R0}\right]\delta n;\\
\nonumber
0&=&\left[-\alpha|\psi_0|^2
+\frac{i\hbar}{2}\gamma_0\frac{\beta\sigma|\psi_0|^2}{(1+\sigma|\psi_0|^2)^2}
\right]a\\
\nonumber
&&+\left[\hbar\omega(1-i\frac{\gamma}{\omega})
-\frac{\hbar^2}{2m}k_{\perp}^2-2\alpha|\psi_0|^2+\frac{i\hbar}{2}Rn_\textrm{R0}-
\frac{i\hbar}{2}\gamma_0(1+\frac{\beta}{1+\sigma|\psi_0|^2}
-\frac{\beta\sigma|\psi_0|^2}{(1+\sigma|\psi_0|^2)^2})
\right]b
+\left[\frac{i\hbar}{2}Rn_\textrm{R0}\right]\delta n;\\
\nonumber
0&=&\left[R|\psi_0|^2\right]a+\left[R|\psi_0|^2\right]b +\left[\gamma+\gamma_\textrm{R}+R|\psi_0|^2\right]\delta n.
\end{eqnarray}
\end{widetext}

The determinant of this system includes the parameter $\gamma$ which we are interested in.
To find $\gamma$, one needs to put the determinant equal to zero and solve the resulting cubic equation on $\gamma$. The analytical solutions are quite cumbersome, therefore we don't present here the final result explicitely. 
The stability conditions are given by the Ljenar-Shipar's criterion. 
If all the three $\gamma$s have negative or zero real part for all $k_{\perp}\in[0,\infty]$ in some range of pumps $P_0$, then the solution is stable in this region. It means that the switching waves are possible between the boundaries of regions corresponding to two stable solutions. The velocity of propagation of the switching wave evidently depends on the pumping intensity. In the case of zero (or close to zero) propagation velocity, a spatial soliton can be formed.



\end{document}